# Low Resistance III-V Heterocontacts to N-Ge

Junkyo Suh[1], Pranav Ramesh[1], Andrew C. Meng[2], Aravindh Kumar[1], Archana Kumar[1], Shashank Gupta[1], Raisul Islam[1], Paul C. McIntyre[2], and Krishna Saraswat[1,2]

[1]Department of Electrical Engineering, Stanford University, Stanford, CA 94305, U.S.A.

[2]Department of Material Science & Engineering, Stanford University, Stanford, CA 94305, U.S.A.

E-mail: {suhjk, saraswat}@stanford.edu

**Abstract:** We experimentally study III-V/Ge heterostructure and demonstrate InGaAs heterocontacts to n-Ge with a wide range of In % and achieve low contact resistivity ($\rho_c$) of $5.0 \times 10^{-8} \Omega cm^2$ for Ge doping of $3 \times 10^{19} cm^{-3}$. This results from re-directing the charge neutrality level (CNL) near the conduction band and benefiting from low effective mass for high electron transmission. For the first time, we observe that the heterointerface presents no temperature dependence despite the two different conduction minimum valley locations of III-V ($\Gamma$-valley) and Ge ($L$-valley), which potentially stems from elastic trap-assisted tunneling through defect states at the interface generated by dislocations. The heterointerface plays a dominant role in the overall $\rho_c$ below $\sim 1 \times 10^{-7} \Omega cm^2$, which can be further improved with large active dopant concentration in Ge by co-doping.

**Introduction:** Ge is a promising material for future CMOS because of its inherently superior carrier mobility and light effective mass. However, large parasitic contact resistance to n-Ge exists due to metal Fermi-level pinning near the valence band [1] and its low dopability (Fig. 1). Attempts to minimize this mainly fall into two categories; 1) interlayer placement between metal and Ge to reduce barrier height [2] and 2) increasing active dopant concentration ($N_d$) by co-doping [3] and/or laser annealing [4]. Laser annealing especially greatly improves dopant activation in Ge above $1 \times 10^{20} cm^{-3}$, but, the benefits vanish during subsequent thermal cycles due to poor thermal stability. On the other hand, the interlayer placement has recently shown great promise in lowering $\rho_c$ to n-Ge by exploiting well-established $Ti(Si_x)/Si:P$ on n-Ge [2].

In this study, we propose III-V as an interlayer and systematically investigate the heterocontact from material and electrical characteristics standpoints. III-V is favorable in that the CNL is located near the conduction band as well as its low effective mass, beneficial for high electron transmission. High In % space was explored (Fig. 2), resulting in nearly zero barrier height [5]. More importantly, we study the temperature dependence of the heterocontact to understand possible carrier transport mechanism at the heterointerface where carriers have to travel across different valleys ($\Gamma$ to $L$ and vice versa). It was found that the heterointerface presents no temperature dependence, which may be indicative of elastic trap-assisted tunneling through defect states from dislocations.

**Experiment:** Ge was epitaxially grown on (100) n-Si [6]. In order to evaluate $\rho_c$ with high precision, MRCTLM structure was fabricated [7]. The key process steps are delineated in Fig. 3. There are two groups of n-Ge doping in terms of $N_d$; 1) in-situ $PH_3$ with $N_d$ of $5 \times 10^{18} cm^{-3}$ during Ge epitaxy and 2) ion implantation (I. I.) with $N_d$ of $3 \times 10^{19} cm^{-3}$. Mo was chosen as contact metal [8] followed by W/Al metallization.

## Results and Discussion

*A. Material Characteristics of III-V Heterocontacts to n-Ge*

3D island growth of InGaAs on Ge was observed with dislocations due to large lattice mismatch (Fig. 4). The surface becomes rougher with In up to 80% and plateaus (Fig. 5). As a result of complete strain relaxation, large FWHM (Fig. 6) and mosaicity (Fig. 7) of III-V were observed by XRD. The estimated threading dislocation density is $\sim 1 \times 10^{10} cm^{-2}$ by Kuhn's model [9] (Fig. 8). HRTEM & EDS analyses were performed, which is consistent with XRD (Fig. 9).

*B. Electrical Characteristics of III-V/n-Ge Contact*

The heterocontact exhibits ohmic behavior and $\rho_c$ was extrapolated from resistance vs. ring spacing plot (Fig. 10). It was found that $\rho_c$ remains high at 60% In regardless of $N_d$ in Ge due to high $R_{contact}$ whereas the effects of the heterointerface ($R_{heterointerface}$) on $\rho_c$ start to manifest above 70% In as a function of $N_d$ in Ge (Fig. 11). It should be pointed out that $\rho_c$ of the group with higher $N_d$ in Ge (I.I.) keeps decreasing with higher In % as opposed to that with lower $N_d$ (in-situ $PH_3$), which means that $R_{heterointerface}$ plays a dominant role in total $\rho_c$ below $1 \times 10^{-7} \Omega cm^2$ and $R_{heterointerface}$ can be engineered by higher $N_d$ in Ge. Low $\rho_c$ of $5.0 \times 10^{-8} \Omega cm^2$ was achieved at InAs/n-Ge with $N_d$ of $3 \times 10^{19} cm^{-3}$, which could be further enhanced with larger $N_d$ by co-doping (Fig. 12). Moreover, optimization of III-V growth temperature is crucial for low $\rho_c$ (Fig. 13).

Understanding of the heterointerface and carrier transport mechanism is critical to be able to engineer $R_{heterointerface}$, and thus the temperature dependence was examined. Since III-V and Ge have different minimum valleys (III-V/Ge systems without crystalline imperfections), carriers have to transit to the other valley across the interface by changing energy and momentum in E-k space, which adds high parasitic $R_{heterointerface}$ [2] because of intervalley scattering which is a function of temperature. However, the heterointerface presents no temperature dependence (Fig. 14), which is possibly attributed to carrier conduction through defect states generated by dislocations [10]. It is plausible that misfit dislocations and/or point defects near the heterointerface lead to energy states in the band-gap [10], which assists carrier transport between $\Gamma$ and $L$ with no need of temperature dependence. Our proposed band [11] and E-k diagrams are illustrated in Fig. 15. Our data may be concrete evidence that the momentum change may occur through defect states.

**Conclusions:** The III-V heterointerface with Ge and its heterocontacts to n-Ge were systematically investigated. Low $\rho_c$ of $5.0 \times 10^{-8} \Omega cm^2$ was achieved, which could be further improved by co-doping (Fig. 16). The heterointerface was found to show no temperature dependence which is suggestive of elastic trap-assisted carrier transport at the interface through defect states.

**Acknowledgement**; This work was supported by Stanford INMP, Global Foundary, TSMC & Lam Research. **References**; [1] T. Nishimura et al., APL, 91 (2007) [2] H. Yu et al., IEDM (2016) [3] B. Yang et al., ISTDM (2012) [4] H. Miyoshi et al., VLSI (2014) [5] Kim et al., APE, 4, (2011) [6] A. Nayfeh et al., APL, 85 (2014) [7] H. Yu et al., EDL, 36 (2015) [8] J. Lin et al., TED, 63 (2016) [9] J. E. Ayers et al., JCG, 135 (1994) [10] R. M. Iutzi, E. A. Fitzgerald, APL, 107 (2015) [11] S. Pal et al., JAC, 645 (2015)

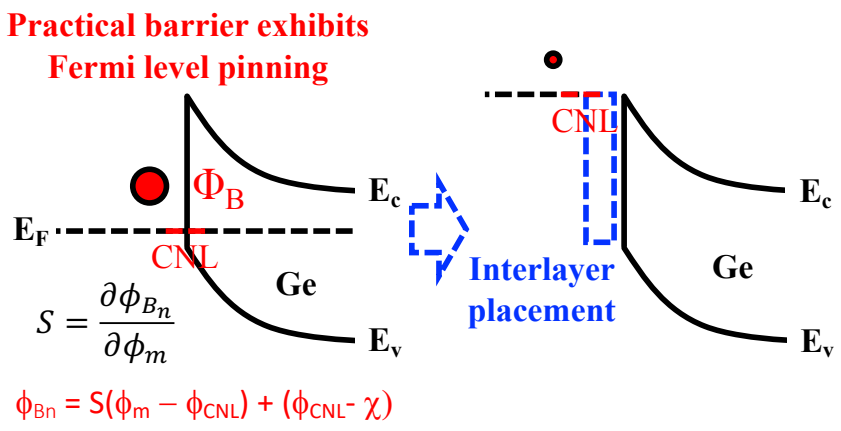

Fig. 1. Barrier height for n-Ge is ~0.6eV (S=~0.05) due to Fermi level pinning [1]. We propose III-V as an interlayer by re-directing CNL and exploiting low effective mass for high electron transmission.

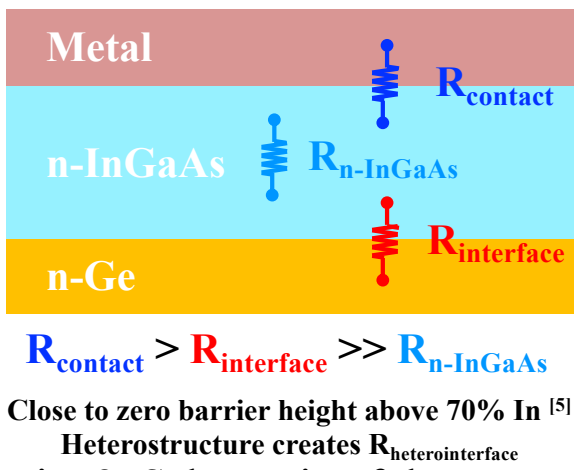

Fig. 2. Schematic of the heterocontact. $R_{heterointerface}$ plays an important role in overall $\rho_c$ at high In %.

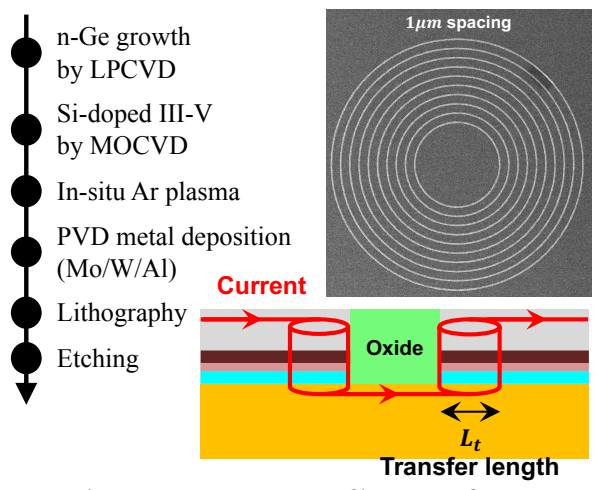

Fig. 3. Process flow of MRCTLM structure [2]. Ge was doped either by in-situ $PH_3$ ($5\times10^{18}$cm$^{-3}$) or P I. I. ($3\times10^{19}$ cm$^{-3}$).

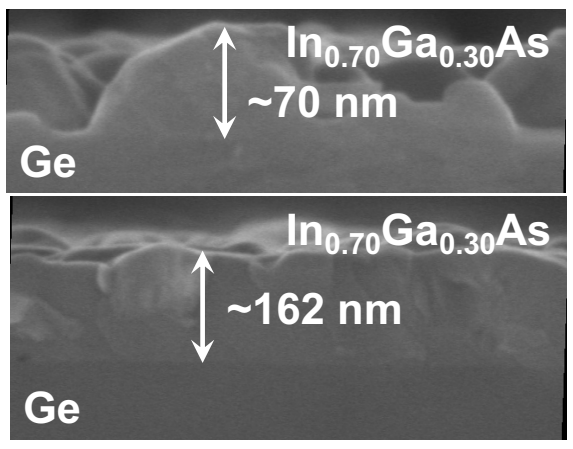

Fig. 4. SEM of III-V/Ge. 3D island growth was observed due to large lattie mismatch.

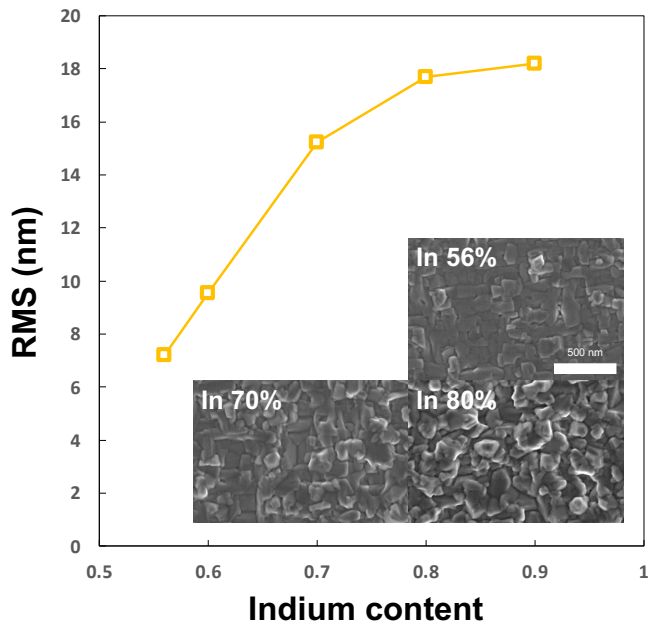

Fig. 5. Surface roughness of III-V/Ge. RMS monotonically increases with In %. In-plane SEM of III-V grown on Ge.

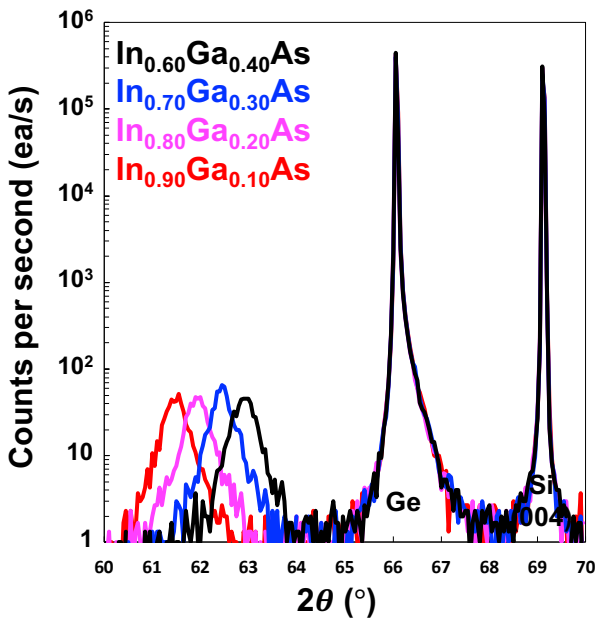

Fig. 6. III-V and Ge were integrated on Si, confirmed by XRD $2\theta-\omega$ scan. Large FWHM of III-V is indicative of defective layers.

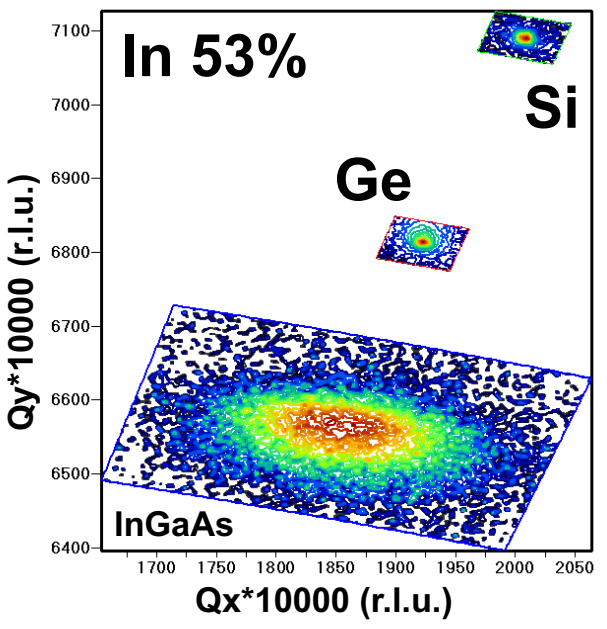

Fig. 7. XRD Reciprocal space mapping. III-V was confirmed to have 0% strain. Large mosaicity due to complete strain relaxation.

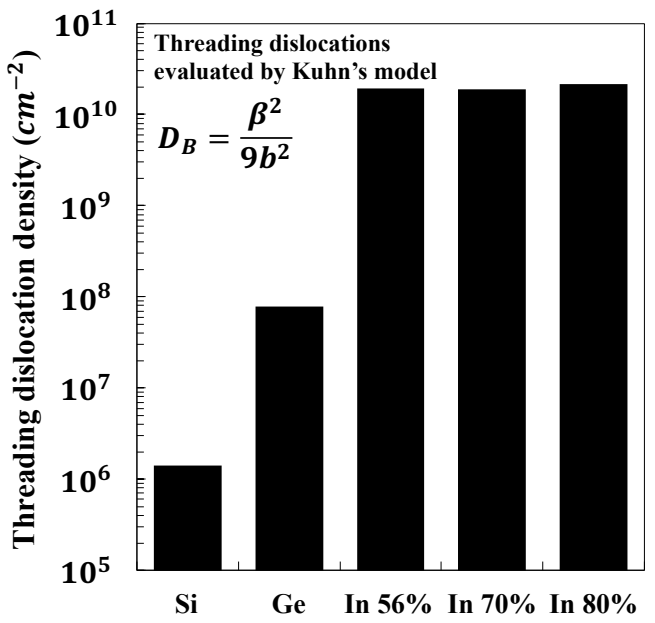

Fig. 8. Estimated dislocation densities of Si, Ge, and III-V from $\omega$ scan. Random distribution and 60° TD in (111) plane were assumed.

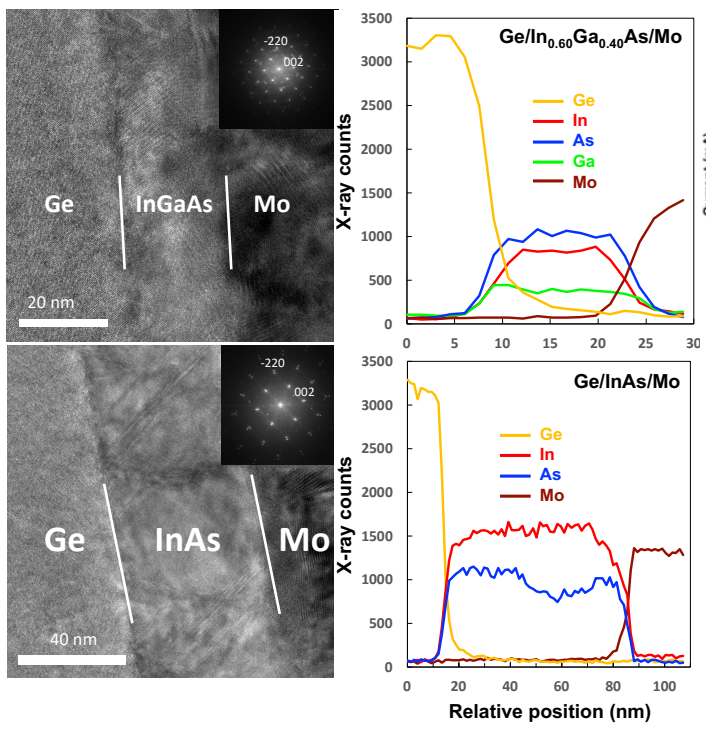

Fig. 9. HRTEM & EDS of 60% In and InAs. Strain relaxation and defects were captured in diffraction pattern. Diffusion of Ge into III-V was observed.

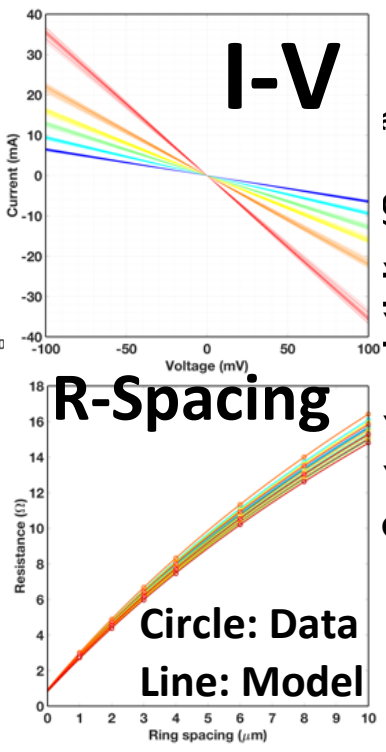

Fig. 10. ohmic behavior of the heterocontact. $\rho_c$ was extracted from resistance vs. spacing.

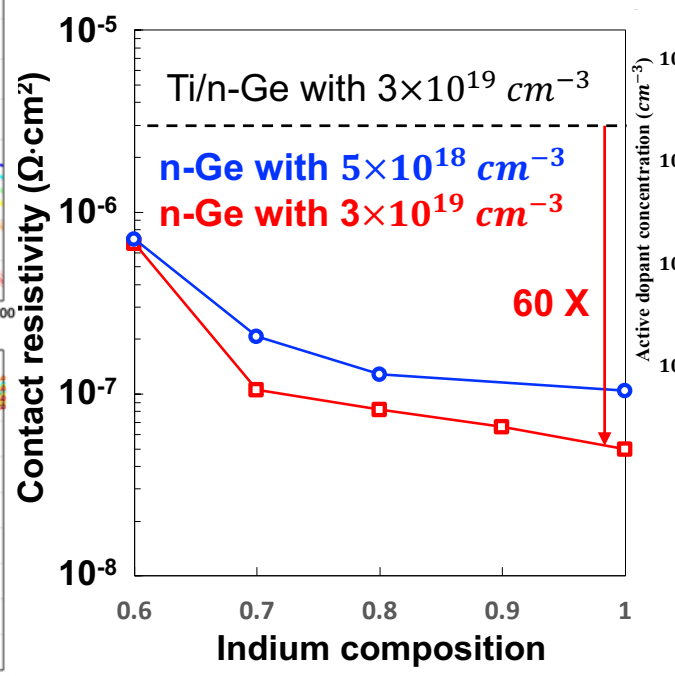

Fig. 11. $\rho_c$ vs. In composition with different $N_d$. The blue line and the red one correspond to Ge with $5\times10^{18}$ cm$^{-3}$ (in-situ $PH_3$) and $3\times10^{19}$ cm$^{-3}$ (I. I.), respectively. 60X improvement was achieved in the heterocontact at $N_d$ of $3\times10^{19}$ cm$^{-3}$.

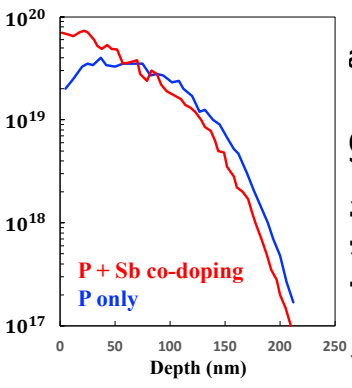

Fig. 12. P+Sb co-doping. $N_d$ > 7e19cm$^{-3}$ was achieved.

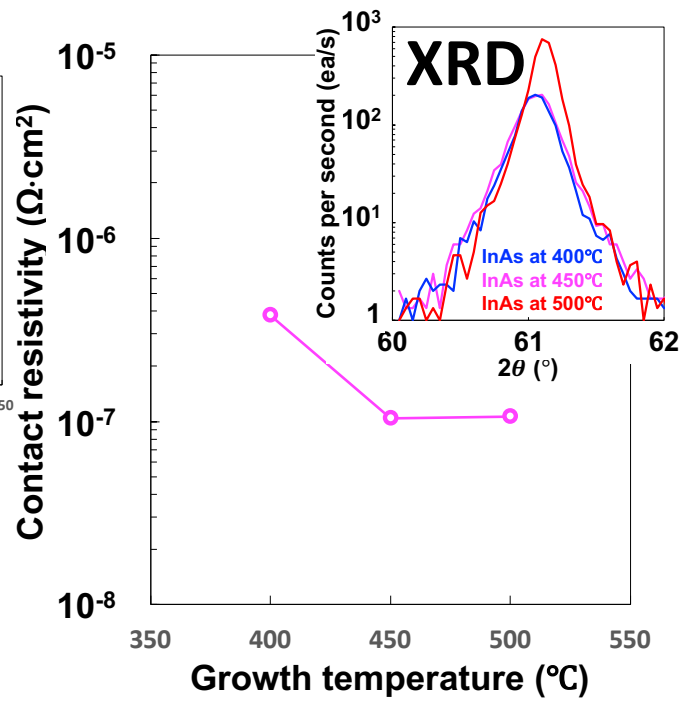

Fig. 13. InAs grown on Ge with $5\times10^{18}$ cm$^{-3}$ at different temperatures. Low temperature growth results in smoother surface, but less $N_d$, thus, $\rho_c$ ↑.

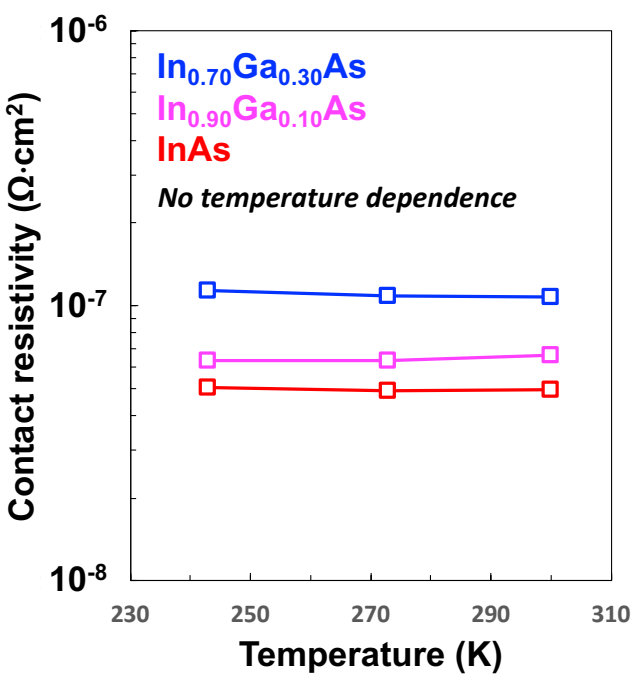

Fig. 14. Temperature dependence of the heterocontacts with various In % showing that $\rho_c$ is almost invariant to temperature.

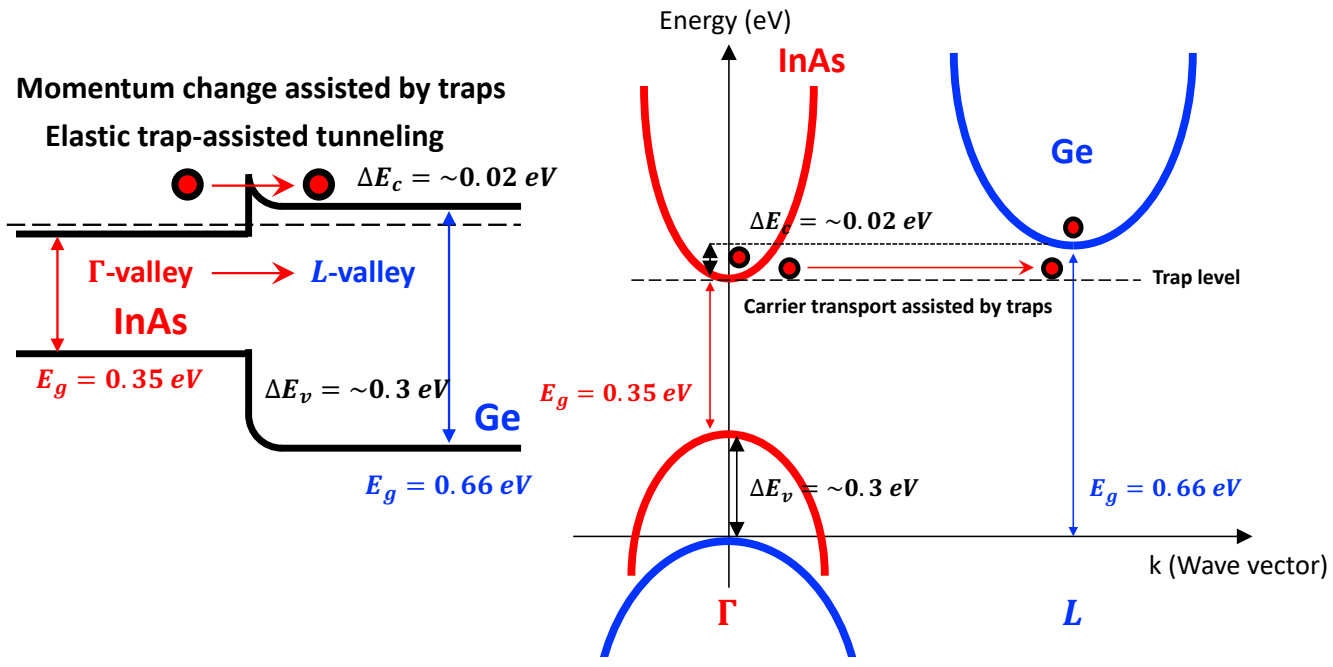

Fig. 15. Band and E-k diagrams of the heterocontact. Conduction band offset (InAs/Ge) is ~0.02eV [11]. Carriers may transit to the other valley at the heterointerface by elastic trap-assist tunneling through defect states, which is temperature independent. Momentum shift is assisted by traps from dislocations.

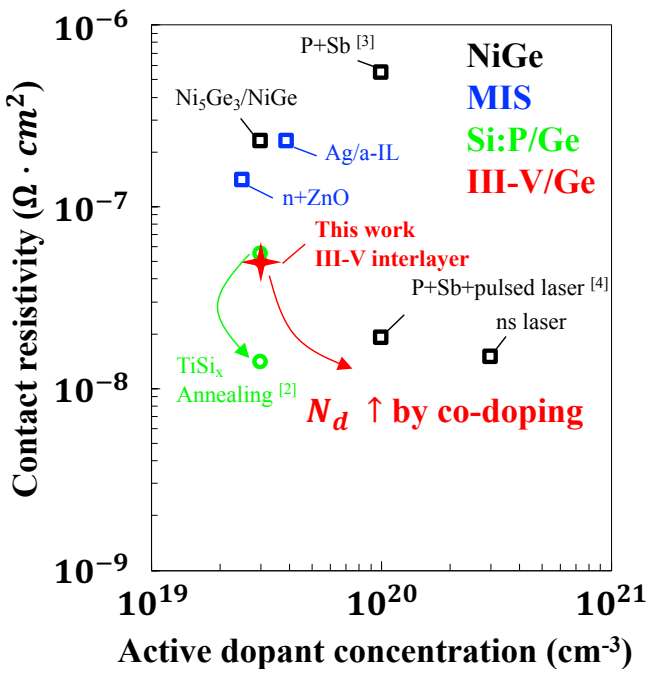

Fig. 16. Benchmark of $\rho_c$ as a function of Ge $N_d$. Co-doping will be deloyed [3].